\documentstyle[preprint,aps,tighten,epsfig]{revtex}

\def\q1{{q^{-1}}}
\def\qq1{{q-q^{-1}}}
\def\djk{{\partial_x^{(q)}}}

\def\nq{{n_{i}}}

\begin{document}
\draft
\title{Thermostatistics of $q$-deformed boson gas}
\author{A. Lavagno $^a$ and P. Narayana Swamy $^b$}
\address{
$^a$ Dipartimento di Fisica and INFN, Politecnico di Torino, Torino, 
I-10129, Italy \\
$^b$ Physics  Department, Southern Illinois University,
Edwardsville, IL 62026, USA}
\maketitle

\begin{abstract}
We show that a natural realization of the thermostatistics of $q$-bosons can be built on the formalism of $q$-calculus and that the entire structure of thermodynamics is preserved if we use an appropriate Jackson derivative in place of the ordinary thermodynamics derivative. This framework allows us to obtain a generalized $q$-boson entropy which depends on the $q$-basic number. We study the ideal $q$-boson gas in the thermodynamic limit which is shown to exhibit Bose-Einstein condensation with higher critical temperature and discontinuous specific heat. 
\end{abstract}

%\vspace{0.5cm}

%\noindent
\pacs{PACS numbers: 05.30.-d; 05.20.-y; 05.70.-a}

\section{Introduction}

The spin-statistics theorem represents one of the fundamental principles of physics and establishes a strict connection between quantum mechanics of many-body systems and quantum statistical mechanics. The complete symmetrization or antisymmetrization of the many body wave function (or the commutation-anticommutation relations in the language of second quantization) reflects the contrasting nature of bosons and fermions. Such quantum many body statistical behavior affects the number of possible states of the system corresponding to the set of occupation numbers and consequently the collective statistical mechanics description.  

The power of the statistical mechanics lies not only in the derivation of the general laws of thermodynamics but also in determining the meaning of all the thermodynamic functions in terms of the microscopic interparticle interaction 
and in providing a collective description of the equilibrium many body system by means of the macroscopic variables such as pressure and internal energy. 

In the recent past there has been increasing emphasis in quantum statistics different from the standard bosons and fermions. Since the pioneering work of Gentile and Green \cite{genti,green}, there have been many extensions beyond the standard statistics  (such as parastatistics, fractional statistics, quon statistics, anyon statistics and quantum groups) which have become  topics of great interest because of the wide range of applications envisaged, from cosmic strings and black holes to the fractional quantum Hall effect and anyonic physics in condensed matter \cite{wil}. 

In the literature there are two principal methods of introducing an intermediate statistical behavior. The first is to deform the quantum algebra of the commutation-anticommutation relations thus deforming the exchange factor between permuted particles. The second method is based on modifying the number of ways of assigning particles to a collection of states and thus the statistical weight of the many-body system. 
The two methods are related but a full connection 
between the quantum mechanics approach and the statistical mechanics approach 
is possible only with the simultaneous knowledge of both. 

One interesting realization of the first approach is the study of exactly solvable statistical systems which has led to a new algebra, the $q$-deformed algebra of creation and annihilation operators, usually called $q$-bosons ($q$-fermions) or $q$-oscillators and related to the general theory of quantum groups \cite{bie,mac}. Many recent investigations in the theory of $q$-bosons have provided much insight into both the mathematical development and the $q$-deformed thermodynamics \cite{lee,su,tus,song,nar,kan,vok,ubri,rodi}. 
However, we believe that a fully consistent formulation connecting the statistical mechanics and the thermodynamics (i.e., thermostatistics) of $q$-bosons has been lacking. 
In particular it is desirable to derive an explicit expression for the entropy of the $q$-bosons, which plays a central role in the thermostatistics of the system and in the information theory. 
It is important to show that the full structure of thermodynamics of $q$-bosons is preserved and the closed loop of thermodynamic relations is satisfied. This is a nontrivial task because there is no {\it a priori} reason that the thermodynamic relations be automatically preserved for the $q$-deformed structures. 

A remarkable example is the Tsallis nonextensive statistics \cite{tsallis}, based on a generalization of the Boltzmann-Gibbs entropy, where the thermodynamic functions such as entropy and internal energy, are deformed but the whole structure of thermodynamics is preserved. Although Tsallis nonextensive thermodynamics is inspired by the (multi)fractal property of a system and does not embody quantum group theory, many papers are devoted to the formal analogies between $q$-oscillators and nonextensive statistics \cite{tsa,abe,joh,ubri2}. The reason for this connection has to do with the common language of the two deformed theories which is the $q$-calculus. 

The $q$-calculus was introduced at the beginning of this century by Jackson \cite{jack} in the study of the basic hypergeometric function and it plays a central role in the representation of the quantum groups \cite{exton}. In fact it has been shown that it is possible to obtain a ``coordinate" realization of the Fock space of the  $q$-oscillators by using the deformed Jackson derivative (JD) \cite{flo,fink}. Moreover we observe that it has recently been shown that the JD can be identified with the generators of fractal and multifractal sets with discrete symmetries \cite{erz}.

Since the thermodynamic functions of nonextensive statistics are deformed by using the framework of $q$-calculus, we expect $q$-calculus to play an important role also in $q$-boson thermostatistics. 

It is the purpose of this paper to show that a fully consistent thermostatistics of $q$-boson gas can be obtained by using an appropriate Jackson derivative rule in the standard thermodynamics relations. In this framework, the whole structure of thermodynamics is preserved and this enables us to derive all the thermodynamic quantities such as the entropy, internal energy and the distribution function in the $q$-deformed theory. Special attention is paid to the study of the ideal $q$-boson gas and the phenomenon of $q$-boson condensation. 

This paper is organized as follows. In Sec. II we review the $q$-boson algebra and outline the modification of the standard boson theory brought about by the $q$-calculus. In Sec. III we determine the distribution function of the $q$-boson gas by utilizing the standard definition of the thermal average of an operator. In Sec. IV we introduce a consistent prescription for the use of the Jackson derivative in the thermodynamic relations. This allows us to obtain in Sec. V the generalized entropy for $q$-bosons and to derive this from the deformed statistical weight. Sec. VI describes the behavior of the ideal $q$-boson gas and the phenomenon of $q$-boson condensation. We report our conclusions in Sec. VII.

\section{$q$-boson algebra and its realizations}

We shall briefly review the theory of  $q$-deformed bosons defined by the $q$-Heisenberg algebra of creation and annihilation operators of bosons introduced by Biedenharn and McFarlane \cite{bie,mac}, derivable through a map from $SU(2)_q$. The 
$q$-boson algebra is determined by the following commutation relations for 
$a$, $a^{\dag}$ and the number operator $N$, thus (for simplicity we omit the particle index)

\begin{equation}
[a,a]=[a^\dag,a^\dag]=0 \; , \ \ \ aa^\dag-q a^\dag a =1 \; , 
\end{equation}
\begin{equation}
[N,a^\dag]= a^\dag \; , \ \ \ [N,a]=-a\; .
\end{equation}

The $q$-Fock space spanned by the orthornormalized eigenstates $\vert n\rangle$
is constructed according to
\begin{equation}
\vert n\rangle=\frac{(a^\dag)^n}{\sqrt{[n]!}} \vert 0\rangle \; , 
\ \ \ a\vert 0\rangle=0 \; ,
\label{fock}
\end{equation}
where the $q$-basic factorial is defined as 
\begin{equation}
[n]!=[n] [n-1] \cdots [1]
\label{brnf}
\end{equation}
and the $q$-basic number $[x]$ is defined in terms of the $q$-deformation parameter 
\begin{equation}
[x]=\frac{q^x-1}{q-1}\; .
\label{bn}
\end{equation}

For the following discussion it is worth observing that the $q$-basic number satisfies the non-additivity property
\begin{equation}
[x+y]=[x]+[y]+(q-1)\, [x]\,[y] \; .
\label{nadd}
\end{equation}
In the limit $q \rightarrow 1$, the $q$-basic number $[x]$ reduces to the ordinary number  $x$ and all the above relations reduce to the standard boson relations.

The actions of $a$, $a^\dag$ on the Fock state $\vert n \rangle$ are given by
\begin{eqnarray}
a^\dag \vert n\rangle &=& [n+1]^{1/2} \vert n+1\rangle\; , \\
a \vert n\rangle&=&[n]^{1/2} \vert n-1\rangle \; ,\\
N \vert n\rangle&=&n\vert n\rangle \; .
\end{eqnarray}
From the above  relations, it follows that
$a^\dag a=[N]$, $aa^\dag=[N+1]$. 

We observe that the Fock space of the $q$-bosons has the same structure as the standard bosons but with the replacement $n!\rightarrow [n]!$ . Moreover the number operator is not $a^\dag a$ but can be expressed as the nonlinear functional relation $N=f(a^\dag a)$ which can be explicitly written formally in the closed form
\begin{equation}
N=\frac{1}{\log q} \log\Big (1+(q-1) a^\dag a \Big )\; .
\label{nop}
\end{equation}

The transformation from Fock observables to the configuration  space (Bargmann holomorphic representation) may be accomplished by choosing \cite{flo,fink}

\begin{equation}
a^\dag=x \; , \ \ \ a=\djk  \; , 
\label{jd}
\end{equation}
where $\djk$ is the Jackson derivative (JD) \cite{jack}
\begin{equation}
\djk f(x)=\frac{f(qx)-f(x)}{x\,(q-1)}\; ,
\end{equation}
which reduces to the ordinary derivative  when $q$ goes to unity and therefore, the JD occurs naturally in $q$-deformed structures \cite{exton}.

\section{Thermal averages and statistical distribution for $q$-boson gas}

Several investigators have studied the equilibrium statistical mechanics of the gas of non-interacting $q$-bosons \cite{lee,su,tus,song,nar,kan,vok,ubri,rodi}. We shall now briefly discuss some of the important results from these studies before introducing our formulation of the thermostatistics of $q$-deformed bosons. 

In the grand canonical ensemble, the Hamiltonian of the non-interacting boson gas is expected to have the following form \cite{lee,su,tus,song}
\begin{equation}
H=\sum_i (\epsilon_i-\mu) \, N_i\; ,
\label{ha}
\end{equation}
where the index $i$ is the state label, $\mu$ is the chemical potential and $\epsilon_i$ is the kinetic energy in the state $i$ with the number operator $N_i$. It should be mentioned that the form of the Hamiltonian is not unique in the literature, where some authors introduce the Hamiltonian which involves the basic number $[N_i]$. The advantage of the form in Eq.(\ref{ha}) is that it describes clearly the number of particles in the spectrum by an integer number  and will allow us to generalize the laws of thermodynamics in a simple manner. 

The thermal average of an operator is written in the standard form
\begin{equation}
\langle {\cal O}\rangle=\frac{ Tr \left ({\cal O} \, e^{-\beta H} \right )}{\cal Z}\; ,
\end{equation}
where $\cal Z$ is the grand canonical partition function defined as 
\begin{equation}
{\cal Z}=Tr \left ( e^{-\beta H} \right )\; ,
\label{pf}
\end{equation}
and $\beta = 1/T$. Henceforward we shall set Boltzmann constant to unity.
Let us observe that the structure of the density matrix $\rho = e^{-\beta H}$ and the thermal average are undeformed. As a consequence, the structure of the partition function is also unchanged. We emphasize that this is not a trivial assumption because its  validity  implicitly amounts to an unmodified structure of the Boltzmann-Gibbs entropy,
\begin{equation}
S=\log W \; ,
\end{equation}
where $W$ stands for the number of states of the system corresponding to the set of occupation numbers $\{ n_i \}$.
Obviously the number $W$ is modified in the $q$-deformed case.
It may be pointed out that in the case of nonextensive $q$-deformed Tsallis statistics, the structure of the entropy is deformed via the logarithm function \cite{tsallis}.

By using the definition in Eq.(\ref{bn}) of the $q$-basic number, the mean value of the occupation number $\nq$ can be calculated starting  from the relation 
\begin{equation}
[n_{i}]=\frac{1}{\cal Z}\, Tr\left ( e^{-\beta H} a^\dag_i a_i\right ),
\end{equation}
and after applying the cyclic property of the trace and using the $q$-boson algebra, it is easy to show that \cite{lee,su,tus}
\begin{equation}
\frac{[n_{i}]}{[n_{i}+1]}=e^{-\beta (\epsilon_i-\mu)} \; .
\label{nqbr}
\end{equation}
The explicit expression for the mean occupation number can be obtained by using the following property of the basic number, 
\begin{equation}
[n_{i}+1]=q\, [n_{i}]+1 \; ,
\label{brnp}
\end{equation}
and hence for $q$ real, 
\begin{equation}
n_{i}=\frac{1}{\log q} 
\log\left (\frac{z^{-1}e^{\beta\epsilon_i}-1}{ z^{-1}e^{\beta\epsilon_i}-q}
\right) \; ,
\label{nqi}
\end{equation}
where $z = e^{\beta \mu}$ is the fugacity. It is easy to see that the above equation reduces to the standard Bose-Einstein distribution when $q\rightarrow 1$. The total number of particles is given by $N=\sum_i\,n_i$.

\section{Jackson derivatives in $q$-thermodynamics relations}

From the definition of the partition function, Eq.(\ref{pf}), and the Hamiltonian, Eq.(\ref{ha}), it follows that the logarithm of the partition function has the same structure as that of the standard boson
\begin{equation}
\log {\cal Z}=-\sum_i \log (1-z e^{-\beta\epsilon_i}) \; .
\end{equation}
This is due to the fact that we have chosen the Hamiltonian to be a linear function of the number operator but it is  not linear in $a^\dag a$ as seen from Eq.(\ref{nop}). For this reason,
the standard  thermodynamic relations in the usual form are ruled out. It is verified, for instance, that 
\begin{equation}
N\ne z \, \frac{\partial}{\partial z} \log {\cal Z}\; .
\end{equation}

As the coordinate space representation of the $q$-boson algebra is realized by the introduction of the JD (see Eq.(\ref{jd})), we stress  that the key point of the $q$-deformed thermostatistics is in the observation that the ordinary thermodynamics derivative with respect to $z$, must be replaced by the JD 
\begin{equation}
\frac{\partial}{\partial z} \Longrightarrow {\cal D}^{(q)}_z \; ,
\end{equation}
where we have defined ${\cal D}^{(q)}_z$ as the Jackson derivative up to a constant (which goes to unity when $q \rightarrow 1$)
\begin{equation}
{\cal D}^{(q)}_z = \frac{q-1}{\log q}\, \partial^{(q)}_z \; .
\end{equation}
Consequently, the number of particles in the $q$-deformed theory can be derived from the relation 
\begin{equation}
N=z \; {\cal D}^{(q)}_z \log {\cal Z}\equiv \sum_i n_i \; ,
\label{num}
\end{equation}
where $n_i$ is the mean occupation number expressed in Eq.(\ref{nqi}).

The usual Leibniz chain rule is ruled out for the JD and therefore derivatives encountered in thermodynamics must be modified according to the following prescription. First we observe that the JD applies only with respect to the variable in the exponential form such as $z=e^{\beta \mu}$ or $y_i=e^{-\beta \epsilon_i}$. Therefore for the $q$-deformed case, any thermodynamic derivative of functions which depend on $z$ or $y_i$ must be converted to derivatives in one of these variables by using the ordinary chain rule and then applying the JD with respect to the exponential variable.  For example, the internal energy in the $q$-deformed case can be written as
\begin{equation}
U=-\left. \frac{\partial}{\partial\beta} \log {\cal Z} \right |_z=\sum_i \frac{\partial y_i}{\partial\beta} \, {\cal D}^{(q)}_{y_i}\log(1-z\,y_i) \; .
\label{int}
\end{equation}
In this case we obtain the correct form of the internal energy
\begin{equation}
U=\sum_i \epsilon_i \, \nq\; ,
\label{un}
\end{equation}
where $n_i$ is the mean occupation number expressed in Eq.(\ref{nqi}).

This prescription is a crucial point of our approach because this allows us to maintain the whole structure of thermodynamics and the validity of the Legendre transformations in a fully consistent manner.

\section{Entropy of the $q$-boson gas and the deformed statistical weight}

In light of the above discussion, we have  the recipe to derive the entropy of the $q$-bosons which leads to  
\begin{eqnarray}
S=-\left. \frac{\partial\Omega}{\partial T}\right |_\mu &&\equiv 
\log {\cal Z} +\beta\sum_i\left.\frac{\partial\kappa_i}{\partial\beta}\right|_\mu  
{\cal D}^{(q)}_{\kappa_i}\log(1-\kappa_i)\nonumber\\
&&=\log {\cal Z} +\beta U-\beta\mu N \; ,
\end{eqnarray}
where $\kappa_i=z\, e^{-\beta\epsilon_i}$,  
$U$ and $N$ are the modified functions expressed in Eqs.(\ref{int}) and (\ref{num}) and $\Omega = - T \log {\cal Z}$ is the thermodynamic potential. 

Using Eqs.(\ref{nqbr})-(\ref{nqi}), after some manipulations, we obtain the entropy as follows 
\begin{equation}
S=\sum_i \Big \{ -\nq \, \log\, [\nq]+(\nq+1)\,  \log\, [\nq+1]-\nq \, \log q
\Big \} \; .
\label{entro}
\end{equation}

The above entropy goes over to the standard boson entropy in the limit $q \rightarrow 1$. It has the compact form which resembles the entropy of the standard boson but with the appearance of the $q$-basic numbers, $[n_i]$ and $[n_i+1]$, in the argument of the logarithmic function and in the presence of the last term, $-\nq\log q$, which follows from non-additivity property of the $q$-basic number. In fact, using Eqs.(\ref{bn}) and (\ref{brnp}) 
can be re-expressed as 
\begin{equation}
\nq\, \log q=\log\, ([\nq+1]-[\nq])\; .
\end{equation}

The expression for the entropy is very relevant to the statistical information about the number of possible states occupied by the $q$-bosons and gives us the desired connection between the deformed quantum algebra and the quantum statistical behavior. It is interesting to observe that in the classical limit, the entropy does not reduce to the standard Boltzmann-Gibbs entropy ($S = - \sum_i n_i \log n_i) $, but remains deformed, except in the limit $q \rightarrow 1$. This result is similar to the case of Greenberg's infinite statistics and the quantum Boltzmann distribution obtained as a particular case of quon statistics \cite{greenberg}. The meaning of this is that the deformation exhibited in the entropy transcends the quantum nature but is built into the theory, somewhat similar to the case of nonextensive Tsallis statistics \cite{tsallis}. The origin of the connection between the two different deformations ($q$-deformed quantum groups and nonextensive statistics) is beyond the scope of this paper and will be reported elsewhere. 

In order to assure consistency, we must now show that the extremization of the entropy with fixed internal energy and number of particles leads to the correct $q$-boson distribution function. The extremum condition can be written as
\begin{equation}
\delta \, \Big ( S-\beta U+\beta\mu N \Big )=0 \; ,
\label{extr}
\end{equation}
where $\beta$ and $\beta\mu$ plays the role of Lagrange multipliers. 

To perform such extremization in the $q$-boson case, we assume that 
the mean occupational number depends on the energy only as a function of $y_i=e^{-\beta\epsilon_i}$, $S=S[n(y_i)]$. Following our prescription described in Sec. IV on the use of JD, 
the above extremization condition can be written as
\begin{equation}
{\cal D}^{(q)}_{y_i} \, \Big ( S-\beta U+\beta\mu N \Big ) \, 
\delta y_i=0\; .
\end{equation}
Employing Eqs. (\ref{num}), (\ref{un}) and (\ref{entro}), and carrying out the JD, the extremization condition reduces to 
\begin{eqnarray}
&&n(q y_i) \left ( \log\frac{[ n(q y_i)+1]}{q\,[n(q y_i)]}-\tilde{\epsilon}_i \right )- 
n(y_i) \left ( \log\frac{[ n(y_i)+1]}{[n(y_i)]}-\tilde{\epsilon}_i \right )+ \nonumber\\
&&n(y_i) \log q- \log\frac{[n(y_i)+1]}{[n(q y_i)+1]}=0 \; ,
\label{nqy}
\end{eqnarray}
where $\tilde{\epsilon}_i=\beta (\epsilon_i-\mu)$. 

The algebraic simplification of the above equation is intractable because of the complexity of the property of $q$-basic numbers. However, it is possible to determine the solution of the equation by observing that for any function $f(x)$, there exists a functional relationship
\begin{equation}
q^{f(x)}=\frac{[f(x)+1]}{[f(qx)+1]} \ \ \Longleftrightarrow \ \ 
\frac{[f(qx)+1]}{[f(qx)]}=\, q \, \frac{[f(x)+1]}{[f(x)]}\; .
\end{equation}
The notation $\Longleftrightarrow $ used here denotes that one relation implies the other and vice versa. The validity of the first relation in the above equation eliminates the last two terms in Eq.(\ref{nqy}) and the validity of the second relation implies that the quantities in parenthesis in  Eq.(\ref{nqy}) are equal, and since $ n(qy_i) \ne  n(y_i)$, for $q \ne 1$, it follows that Eq.(\ref{nqy}) is satisfied if 
\begin{equation}
\frac{[n(y_i)+1]}{[n(y_i)]}=e^{\tilde{\epsilon}_i}\; .
\label{ny34}
\end{equation}

The above relation is equivalent to Eq.(\ref{nqbr}) which implies the mean occupational number $n_i$ of Eq.(\ref{nqi}).

As discussed earlier, the entropy provides the information about the statistical weight $W$ which will be deformed in the case of $q$-boson particles. To investigate this deformation we begin with the basic relation for the entropy
\begin{equation}
S=\log W_q  \; ,
\label{sw}
\end{equation}
where $W_q$ is the deformed statistical weight. Just as the ordinary factorial $n!$ is replaced by the $q$-basic factorial $[n]!$ in the construction $q$-Fock space (see Eq.(\ref{fock})), we assume that this substitution also prevails in the expression for the statistical weight and hence we require
\begin{equation}
W_q=\prod_i \frac{[\nq+g_i-1]!}{[\nq]! \, [g_i-1]!}\; ,
\end{equation}
where $g_i$ denotes the number of subcell levels. 
The reason for this modification lies in the definition of the binomial coefficient in the $q$-combinatorial calculus \cite{exton}. 

Observing that $[n]!$ for large $n$, is given by the  
$q$-Stirling approximation for $q>1$ (see appendix for the explicit derivation)
\begin{equation}
\log\,[n]!\approx n\, \log\,[n]-\frac{n^2}{2} \, \log q\; ,
\end{equation}
the entropy (\ref{sw}) can be written as
\begin{equation}
S=\sum_i \left \{ \nq \, \log\, \frac{[\nq+g_i]}{[\nq]}+g_i \, \log\, \frac{[\nq+g_i]}{[g_i]}-\nq\,g_i\, \log q \right\} \; .
\label{entro2}
\end{equation}

This is similar to the structure of the entropy given by Eq.(\ref{entro}) and therefore the extremization procedure can be carried out as was done before and derive the same condition as in Eq.(\ref{ny34}) except for the factor $g_i$. We observe, however, that the partition operation into subcells is not rigorously true in this context because of the nonextensive property (nonadditivity of the $q$-basic number) of the expression for the entropy in Eq.(\ref{entro}). For this reason, the mean occupation number derived from Eq.(\ref{entro2}) is not rigorously proportional to the factor $g_i$. The nonextensivity implies that the result for the mean occupation number is not entirely independent of the manner in which the energy levels of the particles are grouped into cells.

\section{Ideal $q$-Bose gas and $q$-boson condensation}

We shall now proceed to study the thermodynamic behavior of an ideal $q$-Bose gas and the phenomenon of $q$-boson condensation. For a large volume (and a large number of particles), the sum over all single particle energy states can be transformed to an integral over the energy, as follows
\begin{equation}
\sum_i f(x_i) \ \ \Longrightarrow \ \ 
\frac{2}{\sqrt{\pi}} \, \frac{V}{\lambda^3} \int_0^\infty \!\!\! dx \; x^{1/2} \, f(x)\; ,
\end{equation}
where $x=\beta\epsilon$, $\epsilon = p^2/2m$ is the kinetic energy and 
$\lambda = h/(2\pi m T)^{1/2}$ is the thermal wavelength.

We anticipate that the ground state will be associated with macroscopically large occupation number rather than a zero weight due to $q$-boson condensation. For this reason we need to isolate the ground state and include the contribution from all the other states in the integral. The number density of particles can thus be written as
\begin{equation}
\frac{N}{V}= \frac{2}{\sqrt{\pi}} \, \frac{1}{\lambda^3} \int_0^\infty \!\!\! dx \; x^{1/2} \,  \frac{1}{\log q} \, 
\log \left (\,\frac{z^{-1}e^x-1}{ z^{-1}e^x-q} \right)+\frac{n_0}{V}\; ,
\end{equation}
where $n_0$ is the mean occupational number of the zero momentum state 
\begin{equation}
n_0= \frac{1}{\log q} \, 
\log \left (\,\frac{1-z}{1-q\, z} \right)\; .
\end{equation}

As in the standard boson case, we need to set the range of fugacity $z$ which will correspond to non-negative occupation number. In the case of $q$-bosons we see that the condition is $ z < 1/q$ for $q>1$ and $z<1$ for $q<1$. It should be pointed out that we also have to require the existence of the JD of the mean occupation number which is encountered in the calculation of thermodynamic quantities such as the specific heat and this changes the upper bound of the fugacity $z$. We thus find the correct condition to be $z< z_q$, where we have defined 
\begin{equation}
z_q=\cases{ q^{-2} &if $q>1$ ; \cr 1 &if $q<1$ .\cr} 
\label{zq}
\end{equation}

We will have $q$-boson condensation when the critical combination of density and temperature occurs such that the fugacity will reach its maximum value $z=z_q$. 

Following the prescription of the JD in the $q$-deformed thermodynamics derivatives, we obtain the expression for pressure above the critical point 
\begin{equation}
\left.\frac{P}{T}\right|_>=\frac{1}{\lambda^3} \; g_{_{5/2}} (z,q) \; ,
\label{presa}
\end{equation}
and below the critical point we have 
\begin{equation}
\left.\frac{P}{T}\right|_<=\frac{1}{\lambda^3} \; g_{_{5/2}} (z_q,q) \; .
\label{presb}
\end{equation}
Similar expression can be found for the number of particles above the critical point
\begin{equation}
\left.\frac{N}{V}\right|_>=\frac{1}{\lambda^3} \; g_{_{3/2}} (z,q) \; ,
\label{totnpa}
\end{equation}
and below the critical point we have
\begin{equation}
\left.\frac{N}{V}\right|_<=\frac{n_0}{V}+
\frac{1}{\lambda^3} \; g_{_{3/2}} (z_q,q) \; .
\label{totnpb}
\end{equation}

In the above equations we have defined the $q$-deformed $g_{_n}(z,q)$ functions as
\begin{eqnarray}
g_{_n}(z,q)&=&\frac{1}{\Gamma (n)} \int_0^\infty \!\!\! dx \; x^{n-1} 
\frac{1}{\log q} \log\left (\,\frac{z^{-1}e^x-1}{ z^{-1}e^x-q} \right) \nonumber\\
&\equiv& \frac{1}{\log q} \left ( 
\sum_{k=1}^{\infty} \frac{(zq)^k}{k^{n+1}} - 
\sum_{k=1}^{\infty} \frac{z^k}{k^{n+1}}   \right )  \; .
\label{gn}
\end{eqnarray}

In the limit $q\rightarrow 1$, the deformed $g_n(z,q)$ functions reduce to the standard $g_n(z)$. In Fig. 1 and 2 we present the behavior of $g_{_{3/2}}(z,q)$ and $g_{_{5/2}}(z,q)$ as a function of $z$ for different values of the parameter $q$. 

The internal energy can be calculated considering the thermodynamic limit of Eq.(\ref{int}) by means of the JD recipe. Using the expression for the pressure,  Eq.(\ref{presa}), it is easy to verify that as in the undeformed case, 
the following well-known relation is satisfied for the $q$-bosons,
\begin{equation}
U=\frac{3}{2} \, PV \; .
\end{equation}

We can calculate the critical temperature by using the same method as in the standard boson case. Comparing the ratio of the critical temperature $T_c^q$ of the $q$-deformed gas with that of the standard boson $T_c$ at the same density, we find
\begin{equation}
\frac{T_c^q}{T_c}=\left ( \frac{g_{_{3/2}}(1)}{g_{_{3/2}}(z_q,q)} \right )^{2/3} \; ,
\end{equation}
where $g_{_{3/2}}(1)=2.61$ is the value of the undeformed function when $z=1$.
In Fig. 3 we show the plot of the above ratio as a function $q$. We observe that the critical temperature of the $q$-boson is always higher than the standard boson and for $q>1$ there is a rapid increase of the critical temperature $T^q_c$ for small values of $q$. For example, for $q=1.01$, $T^q_c$ increases by $18\%$ and for $q=1.1$, $T^q_c$ increases by $75\%$ with respect to the standard value. 

Applying the thermodynamic limit to the entropy of the $q$-boson in Eq.(\ref{entro}), we obtain the entropy per unit volume above the critical point with a structure similar to that of the standard boson,
\begin{equation}
\left.\frac{S}{V}\right|_>=\frac{1}{\lambda^3} \left(\,\frac{5}{2} \; g_{_{5/2}} (z,q)- g_{_{3/2}} (z,q) \log z \right) \; ,
\label{entroa}
\end{equation}
and below the critical point
\begin{equation}
\left.\frac{S}{V}\right|_<= \frac{5}{2}\, \frac{1}{\lambda^3}\, g_{_{5/2}} (z_q,q) \; .
\label{entrob}
\end{equation}

Let us observe that the generalized $q$-boson entropy obeys the third law of thermodynamics. In fact, in Eq.(\ref{entrob}), $g_{_{5/2}} (z_q,q)$ has a finite value that depends on $q$ and the entropy approaches zero in the limit of zero temperature.

As in the ordinary Bose condensation it is possible to show that in the $q$-boson condensation also a Clausius-Clapeyron equation holds and first order phase transition occurs. In fact it is easy to see that below the critical point the following equation is satisfied 
\begin{equation}
\left.\frac{dP}{dT}\right|_<=\frac{L_q}{T \, v_c} \; ,
\end{equation}
where $v_c$ is the critical specific volume, defined as
\begin{equation}
v_c=\frac{\lambda^3}{ g_{_{3/2}}(z_q,q)}\; ,
\end{equation}
$L_q$ is the q-deformed latent heat given by 
\begin{equation}
L_q=T \, \Delta s= \frac{5}{2} \; T \; \frac{g_{_{5/2}} (z_q,q)}{g_{_{3/2}}(z_q,q)} \; ,
\end{equation}
and where $\Delta s$ is the difference in specific entropy across the transition region.

We now proceed to calculate the heat capacity of the $q$-boson gas, starting from the thermodynamic definition 
\begin{equation}
C_v=\left. \frac{\partial U}{\partial T}\right|_{V,N}\; .
\label{cvt}
\end{equation}

For this purpose we first need the derivative of the fugacity with respect to $T$ (or $\beta$), keeping $V$ and $N$ constant.  To apply the JD prescription described  earlier in Sec. IV, we start from the expression for the total number of particles, Eq.(\ref{num}) and the identity (since the number of particles is kept constant) 
\begin{equation}
\frac{\partial}{\partial \beta}\sum_i \log \left ( \frac{1- \kappa_i}{1-q \kappa_i}  \right ) = 0 \; ,
\end{equation}
where $\kappa_i= z \, e^{- \beta \epsilon_i}$. This identity can be rewritten according to our JD recipe as
\begin{equation}
\sum_i \, \frac{\partial \kappa_i}{\partial \beta}\, {\cal D}_{\kappa_i}^{(q)}\log 
\left ( \frac{1- \kappa_i}{1-q \kappa_i}  \right ) = 0\; ,
\end{equation}
and now evaluating in the limit $V\rightarrow\infty$, we obtain
\begin{equation}
\left. \frac{1}{z}\, \frac{\partial z}{\partial\beta}\right|_{V,N}=\frac{3}{2}\;  \frac{1}{\beta} \; \frac{{\cal D}^{(q)}_z g_{_{5/2}} (z,q)}{ {\cal D}^{(q)}_z g_{_{3/2}} (z,q)} \; .
\label{dzb}
\end{equation}

We shall now proceed to calculate the heat capacity. Using the discrete expression of internal energy (\ref{int}), Eq.(\ref{cvt}) can be expressed as
\begin{equation}
C_v=-\beta^2 \sum_i \, \epsilon_i\,\frac{\partial \kappa_i}{\partial \beta}\; \frac{1}{\log q} {\cal D}_{\kappa_i}^{(q)}\log 
\left ( \frac{1- \kappa_i}{1-q \kappa_i}  \right ) \; .
\end{equation}
Carrying out the limit $V\rightarrow\infty$ and utilizing Eqs.(\ref{totnpa}) and (\ref{dzb}), we obtain the following expression for the specific heat per particle above the critical point
\begin{equation}
\left. \frac{C_v}{N}\right|_>= \frac{15}{4}\; \frac{z\, {\cal D}^{(q)}_z g_{_{7/2}} (z,q)}{ g_{_{3/2}} (z,q)}-\frac{9}{4}\; \frac{z\, {\cal D}^{(q)}_z g_{_{5/2}} (z,q)}{ g_{_{3/2}} (z,q)}\; \frac{{\cal D}^{(q)}_z g_{_{5/2}} (z,q)}{ {\cal D}^{(q)}_z g_{_{3/2}} (z,q)} \; ,
\label{cva}
\end{equation}
and similarly the specific heat below the critical point
\begin{equation}
\left. \frac{C_v}{N}\right|_<=\frac{15}{4}\; \frac{z\, {\cal D}^{(q)}_z g_{_{7/2}} (z,q)|_{z=z_q} }{\lambda^3/v}\; ,
\label{cvb}
\end{equation}
where $v$ is the specific volume below the critical temperature that can be expressed, by means of Eq.(\ref{totnpb}), in terms of the critical temperature $T_c$ as follows
\begin{equation}
\frac{v}{\lambda^3}=\frac{1}{ g_{_{3/2}} (z,q) }\, \left ( \frac{T}{T^q_c}\right )^{3/2} \; .
\end{equation}

The above expressions have the same structure as that of the undeformed boson but the difference arises from the property 
$z \, {\cal D}^{(q)}_z g_{_n}(z,q)\ne g_{_{n-1}}(z,q)$, 
where the equality is true for ordinary derivative only. From this observation it is easy to see that in the limit $q\rightarrow 1$ the specific heat reduces to the well-known undeformed result. 

As usual, the classical limit can be achieved considering the limit $z\rightarrow 0$. In this limit the deformed $g_{_n}(z,q)$ functions reduce to 
\begin{equation}
g_n(z,q)\rightarrow \frac{q-1}{\log q} \; z \; ,
\end{equation}
and from Eq.(\ref{cva}), the ``classical" limit of the specific heat per particle number reduces to 
\begin{equation}
\left.\frac{C_v}{N}\right|_{cl}=\frac{3}{2} \; \frac{q-1}{\log q} \; .
\end{equation}
As discussed before in the context of the entropy, the $q$-deformation persists also in the classical limit. 

We expect small deviations from undeformed behavior in the experimental observables, therefore only values of $q$ close to standard value $q=1$ are physically significant. Small deformation leads to a negligible departure from the high temperature limit of the specific heat but implies sharp deformation of the behavior of the specific heat in the range of the critical temperature. 
To exemplify this feature we plot in Fig. 4 the specific heat as a function of $T/T_c^q$ for $q=1.05$. We have chosen a value of $q$ in the range $q>1$ because this region appears particularly interesting with a higher critical temperature for small $q$ (see Fig. 3). For this value of $q$ the critical temperature $T^q_c$ is increased by $48\%$ relative to the standard boson case. 
We observe that for $q\ne 1$, the specific heat shows a discontinuous $\lambda$ point behavior. This is a characteristic of $q$-deformation as has been observed in other investigations \cite{ubri,rodi}. 

Using Eqs.(\ref{cva}) and (\ref{cvb}) we can calculate the jump in the specific heat $\Delta (C_v/N)$ at the critical temperature as a function of $q$. In Fig. 5 we plot this behavior. We observe that the jump is an increasing function of $q$. 

Although the model that we have investigated is based on the Hamiltonian of noninteracting particles, we note that the jump in the specific heat is of the order of the experimental value in the case of Bose condensation in $^{87}$Rb atoms \cite{ensher}.

\section{Conclusion}
The outstanding problem in the theory of $q$-bosons has been the lack of a  demonstration that the thermodynamic relations follow from the $q$-calculus framework. 

In this paper, we have shown that the whole structure of thermodynamics is preserved if the ordinary derivatives are replaced by the Jackson derivatives following the prescription described in Sec. IV.  We establish a fully consistent set of relations between the thermodynamic functions (partition function, internal energy, mean occupation number) and this enables us to derive the entropy of $q$-bosons. The $q$-deformed entropy so obtained has been shown to follow from the deformed statistical weight as a consequence of the $q$-combinatorial calculus known in the literature \cite{exton}. This result represents a close connection between the quantum deformed algebra and the quantum statistical approach.

The expression for the entropy is nonextensive because of the non-additive property (\ref{nadd}) of the $q$-basic numbers. We find that for $q \ne 1$, the entropy remains deformed in the classical limit as is also true of the other thermodynamic functions. This can be understood by observing that the deformation arises from quantum groups but the nature of the deformation is inherently contained in the theory. A similar feature is found in the nonextensive Tsallis statistics and infinite statistics where the deformation persists in the classical limit \cite{tsallis,greenberg}. 

In this framework, we have studied the basic properties of the ideal $q$-Bose gas in the thermodynamic limit and the phenomenon of $q$-boson condensation. We find that the critical temperature of the $q$-boson is always higher than that of the standard boson. The behavior of the specific heat exhibits a discontinuity at the transition point, which is in qualitative agreement, for values of $q$ close to unity, with experimental data in the case of a dilute gas of Rubidium atoms \cite{ensher}. 

We observe that for an ideal Bose gase the specific heat is continuous. On the basis of the Ginsburg-Landau theory of $\lambda$-points, a discontinuous behavior of the specific heat implies a broken symmetry in the transition characterized by an order parameter. The deformation of the algebra in $q$-boson theory implies a broken permutation symmetry of the standard boson wave function. Therefore, the recent experimental data \cite{ensher} can be interpreted as an indication of the effects due to $q$-deformation in Bose-Einstein condensation, where the order parameter of the phase transition depends on $q$. 

Although we employed the non-symmetric $q$-deformation in this investigation, all the results can be easily extended by using the symmetric $q$-calculus $(q \leftrightarrow q^{-1})$. We have confined our study to the $q$-deformation of bosons. It may be worthwhile to investigate the theory of $q$-fermions in this framework.

Our theoretical framework and the results appear to provide a deeper insight into the behavior of the $q$-boson gas. We believe that the results derived here may be relevant to future investigations, and may be of interest from theoretical as well as experimental point of view.

\vspace{.2in}

\noindent
{\bf Acknowledgments}
\vspace{.1 in}

We are grateful to P. Quarati for encouragement and useful discussions. One of us (A.L.) would like to thank the Physics Department of Southern Illinois University for warm hospitality where this work was done.

\appendix
\section{}

Here we present a derivation for the approximation of $q$-basic factorial, $[n]!$ for large $n$, which is the analog of the Stirling approximation. This is employed in the derivation of the entropy in Sec. V. We limit our discussion to the case of $q>1$. 

Starting from the definition (\ref{brnf}) of the $[x]!$ and using the property (\ref{brnp}) of the $q$-basic number, it is possible write any product factor contained in the $[x]!$ as follows:

\begin{eqnarray}
&&[n] \ \ \ \ \ \, = \ \ \ \ \, [n] \nonumber\\
&&[n-1]=q^{-1} \, [n] -q^{-1}\nonumber\\
&&[n-2]=q^{-2} \, [n] -q^{-2}-q^{-1}\nonumber\\
&&[n-3]=q^{-3} \, [n] -q^{-3}-q^{-2}-q^{-1}\nonumber\\
&&\ \ \ \ \vdots \nonumber\\
&&[n-k]=q^{-k} \, [n] -q^{-k}-q^{-k+1}- \cdots -q^{-1}\nonumber\\
&&\ \ \ \ \vdots \nonumber\\
&&[1] \ \ \ \ \ \, =q^{-n+1} \, [n] -q^{-n+1}-q^{-n+2}- \cdots -q^{-1}\; .
\label{appefat}
\end{eqnarray}
For $n\gg 1$, the leading term is seen to be 
\begin{equation}
[n]!\approx \, [n]^n \; q^{-\sum_{k=0}^{n-1} {\displaystyle k}} \; 
\left ( 1-\frac{n}{q^n}\right )\; ,
\end{equation}
where the first term arises from the product of the first term in each of the equations (\ref{appefat}) and the second term is the result of the sum of the dominant corrections. 

The above equation can be rewritten as
\begin{equation}
[n]!\approx \, [n]^n \; q^{-n(n-1)/2} \left ( 1-\frac{n}{q^n}\right )\; .
\end{equation}
Taking the logarithm on both sides, we have 
\begin{equation}
\log[n]!\approx n\, \log[n]-\frac{n^2}{2} \, \log q -\frac{n}{q^n}\; ,
\label{stir}
\end{equation}
where we observe that the last term, which follows from the approximation:  $\log(1-n/q^n)\approx -n/q^n$, is very small and significant only for $q$ very close to unity ($\vert q-1\vert <10^{-3}$). We neglect this term in the derivation of the entropy in Sec. V.

We have verified numerically that the derived $q$-Stirling approximation 
is very good for large $n$. For example, for $q=1.5$ it is correct to an error of $1.8\%$ for $n=100$, $0.2\%$ for $n=1000$ and $0.04\%$ for $n=5000$.

\begin{figure}[htb]
\mbox{\epsfig{file=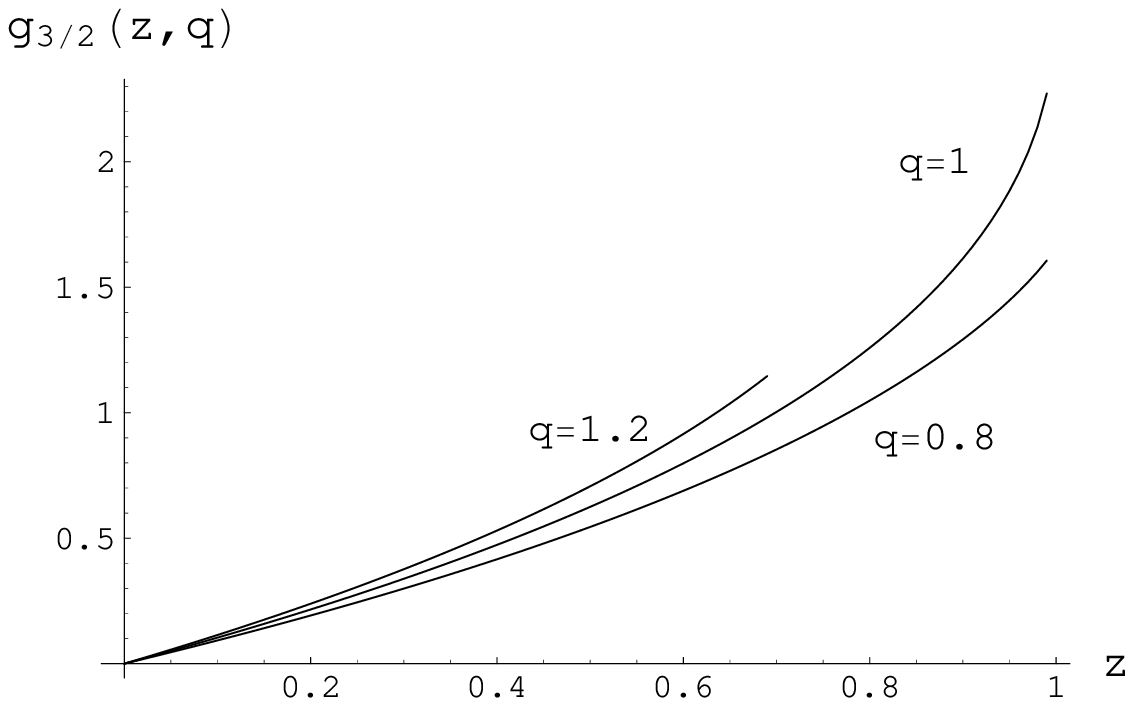,width=0.95\textwidth}}
\caption[]{The behavior of $g_{_{3/2}} (z,q)$ as a function of $z$ for different values of $q$. The value $q=1$ corresponds to the standard $g_{_{3/2}}(z)$ boson function. For $q>1$ the upper bound of $z$ is $1/q^2$ and for $q<1$ it is unity 
(see Eq.(\ref{zq})) 
due to the existence of the JD of the $g_{_{n}} (z,q)$.}
\end{figure}

\begin{figure}[htb]
\mbox{\epsfig{file=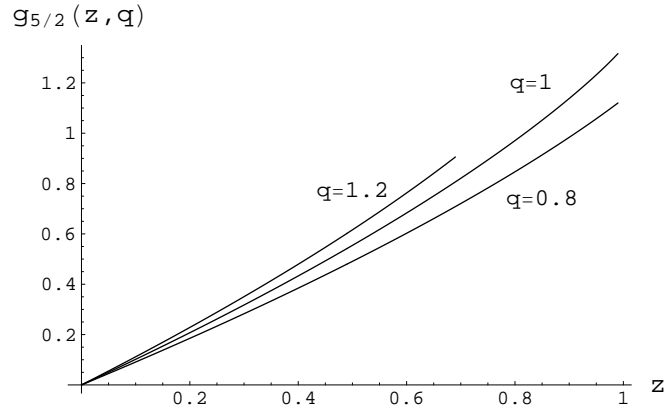,width=0.95\textwidth}}
\caption[]{Same as Fig. 1 for the function $g_{_{5/2}} (z,q)$.}
\end{figure}

\begin{figure}[htb]
\mbox{\epsfig{file=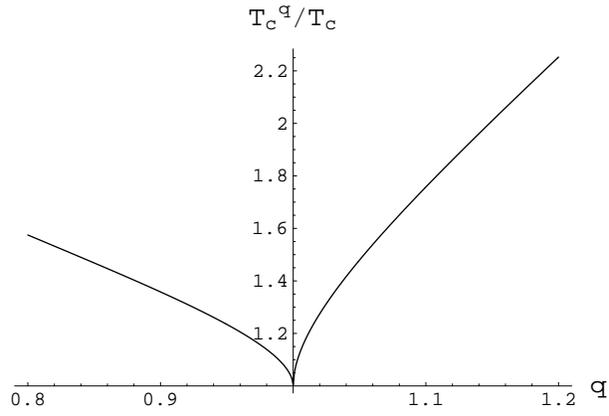,width=0.95\textwidth}}
\caption[]{The ratio $T^q_c/T_c$ of the deformed critical temperature $T^q_c$ and the undeformed ($q=1$) $T_c$ as a function of $q$.}
\end{figure}

\begin{figure}[htb]
\mbox{\epsfig{file=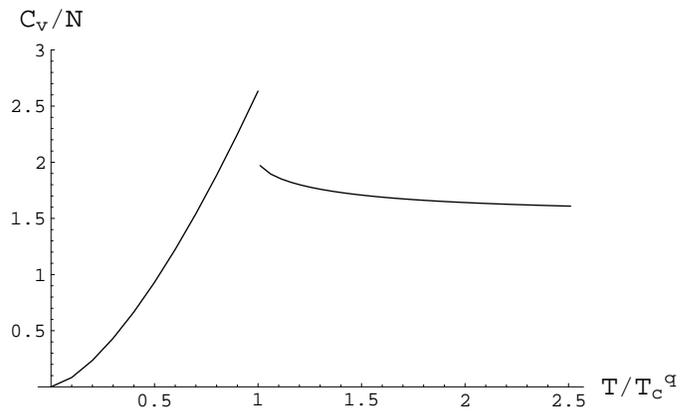,width=0.95\textwidth}}
\caption[]{The specific heat $C_v/N$ as a function of $T/T^q_c$ for $q=1.05$.}
\end{figure}

\begin{figure}[htb]
\mbox{\epsfig{file=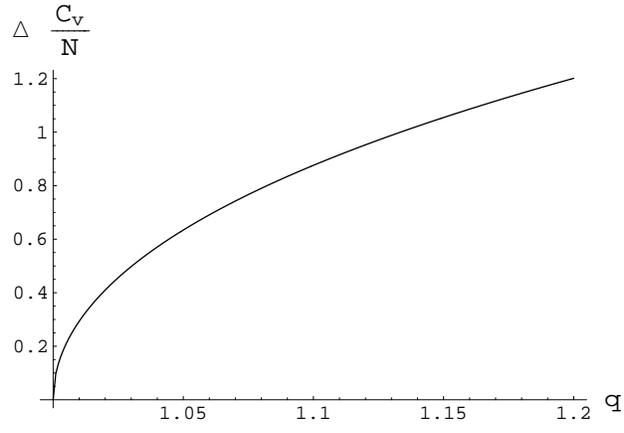,width=0.95\textwidth}}
\caption[]{The jump in the specific heat $\Delta (C_v/N)$ at the critical temperature $T^q_c$ as a function of $q$.}
\end{figure}

\end{document}